# Satellite-Based Entanglement Distribution Over 1200 kilometers

Juan Yin[1,2], Yuan Cao[1,2], Yu-Huai Li[1,2], Sheng-Kai Liao[1,2], Liang Zhang[2,3], Ji-Gang Ren[1,2], Wen-Qi Cai[1,2], Wei-Yue Liu[1,2], Bo Li[1,2], Hui Dai[1,2], Guang-Bing Li[1,2], Qi-Ming Lu[1,2], Yun-Hong Gong[1,2], Yu Xu[1,2], Shuang-Lin Li[1,2], Feng-Zhi Li[1,2], Ya-Yun Yin[1,2], Zi-Qing Jiang[3], Ming Li[3], Jian-Jun Jia[3], Ge Ren[4], Dong He[4], Yi-Lin Zhou[5], Xiao-Xiang Zhang[6], Na Wang[7], Xiang Chang[8], Zhen-Cai Zhu[5], Nai-Le Liu[1,2], Yu-Ao Chen[1,2], Chao-Yang Lu[1,2], Rong Shu[2,3], Cheng-Zhi Peng[1,2], Jian-Yu Wang[2,3], Jian-Wei Pan[1,2]

[1]Shanghai Branch, Department of Modern Physics and National Laboratory for Physical Sciences at the Microscale, University of Science and Technology of China, Shanghai 201315, China
[2]CAS Center for Excellence and Synergetic Innovation Center in Quantum Information and Quantum Physics, University of Science and Technology of China, Shanghai 201315, China
[3]Key Laboratory of Space Active Opto-Electronic Technology, Shanghai Institute of Technical Physics, Chinese Academy of Sciences, Shanghai 200083, China
[4]The Institute of Optics and Electronics, Chinese Academy of Sciences, Chengdu 610209, China
[5]Shanghai Engineering Center for Microsatellites, Shanghai 201203, China
[6]Key Laboratory of Space Object and Debris Observation, Purple Mountain Observatory, Chinese Academy of Sciences, Nanjing 210008, China
[7]Xinjiang Astronomical Observatory, 150, Science-1 Street, Urumqi 830011, China
[8]Yunnan Observatories, Chinese Academy of Sciences, Kunming 650011, China

**Abstract:**

**Long-distance entanglement distribution is essential both for foundational tests of quantum physics and scalable quantum networks. Owing to channel loss, however, the previously achieved distance was limited to ~100 km. Here, we demonstrate satellite-based distribution of entangled photon pairs to two locations separated by 1203 km on the Earth, through satellite-to-ground two-downlink with a sum of length varies from 1600 km to 2400 km. We observe a survival of two-photon entanglement and a violation of Bell inequality by $2.37 \pm 0.09$ under strict Einstein locality conditions. The obtained effective link efficiency at 1200 km in this work is over 12 orders of magnitude higher than the direct bidirectional transmission of the two photons through the best commercial telecommunication fibers with a loss of 0.16 dB/km.**

Quantum entanglement, first recognized by Einstein, Podolsky and Roson (*1*), and by Schrodinger (*2*), describes a physical phenomenon that the quantum states of a many-particle system cannot be factorized into a product of single-particle wave-functions, even when these particles are separated by large distances. Entangled states have been produced in laboratories (*3-5*) and exploited to test the contradiction between classical local hidden variable theory and quantum mechanics using Bell's inequality (*6*). It is of fundamental interest to distribute entangled particles over increasingly large distance and study the behavior of entanglement at extreme conditions. Practically, large-scale dissemination of entanglement—eventually at a global scale—is useful as the essential physical resources for quantum information protocols such as quantum cryptography (*7*), quantum teleportation (*8*) and quantum networks (*9*).

However, so far entanglement distribution has only been achieved at a physical separation up to ~100 km (*10*). This is mainly limited by the photon loss in the channel (optical fibers or terrestrial free space), which normally scales exponentially with the channel length. For example, via bidirectional distribution of an entangled source of photon pairs with 10-MHz count rate directly through two 600-km telecommunication fibers with a loss of 0.2 dB/km, eventually one would only obtain $10^{-17}$/s two-photon coincidence events. When the transmitted photons are attenuated to a level comparable to the dark counts of the single-photon detectors, the entanglement can't be established owing to the low signal-to-noise ratio. To improve the signal-to-noise ratio, one cannot noiselessly amplify the entangled photons in the channel due to quantum non-cloning theorem (*11*), but has to find radically new methods to reduce the link attenuation.

To this end, one solution is the protocol of quantum repeaters (*12*) that divide the whole transmission line into $N$ smaller segments, and combine the functionalities of entanglement swapping (*13*), entanglement purification (*14*) and quantum storage (*15*). There has been considerable progress in the demonstrations of these building blocks (*16-18*) and proof-of-principle quantum repeater nodes (*19,20*). However, the practical usefulness of the quantum repeaters is still hindered by the challenges of simultaneously realizing and integrating all the key capabilities, including most importantly long storage time and high retrieval efficiency (*21*).

A unique shortcut to global-scale quantum networks is making use of satellite and space-based technologies. A satellite can conveniently cover two distant locations on the Earth separated by thousands of kilometers. The key advantage of this approach is that the photon loss and turbulence predominantly occurs in the lower ~10 km of the atmosphere, and most of the photons' transmission path is virtually in vacuum with almost zero absorption and decoherence. Previously, ground-based feasibility studies have demonstrated bidirectional distribution of entangled photon pairs through two-link terrestrial free-space channel—with violations of Bell inequality—over a distance 600 m (*22*), 13 km that goes beyond the effective atmospheric thickness (*23*), and 102 km with ~80 dB effective channel loss comparable to satellite-to-ground two downlink (*10*). In addition, quantum communications on moving platforms in a high-loss regime and under turbulent conditions were also tested (*24,25*). After these feasibility studies, a sophisticated satellite dedicated for quantum science experiments, weighted 635 kg and named *Micius* (SOM), has been developed and successfully launched from Jiuquan, China, to an altitude of ~500 km.

For the mission of entanglement distribution, three ground stations are cooperating with the satellite, located in Delingha (37°22'44.43''N, 97°43'37.01E; altitude 3,153 m) in Qinghai, Nanshan (43°28'31.66''N, 87°10'36.07''; altitude 2,028 m) in Urumqi, Xinjiang, and Gaomeigu Observatory (26°41'38.15''N, 100°1'45.55''E; altitude 3,233 m) in Lijiang, Yunnan. The physical distance between Delingha and Lijiang (Nanshan) is 1203 km (1120 km). The separation between the orbiting satellite and these ground stations varies from 500 km to 2000 km. The effective laboratory space is thus greatly increased and provides a new platform for quantum networks as well as for probing the validity of quantum mechanics.

Here, we demonstrate bidirectional distribution of entangled photon pairs through satellite-to-ground two-downlink channels. By developing ultrabright spaceborne two-photon entanglement sources and high-precision acquiring, pointing and tracking (APT) technology, we establish entanglement between two single photons separated over 1203 km, with an average two-photon count rate of 1.1 Hz, and state fidelity of $0.869 \pm 0.085$. Using the distributed entangled photons, we perform Bell test at space-like separation and without the locality and the freedom-of-choice loophole.

A schematic layout of our experimental set-up is illustrated in Fig. 1. We start by describing our design of spaceborne source of entangled photon pairs. As shown in Fig. 2A, a continuous-wave laser diode with a central wavelength of 405 nm and a linewidth of ~160 MHz is used to pump a periodically poled KTiOPO$_4$ (PPKTP) crystal inside a Sagnac interferometer. The pump laser, split by a polarizing beam splitter (PBS), passes through the nonlinear crystal in clockwise and anticlockwise direction simultaneously, which produces down-converted photon pairs at ~810 nm wavelength in polarization-entangled states close to the form $|\psi\rangle_{12} = (|H\rangle_1 |V\rangle_2 + |V\rangle_1 |H\rangle_2)/\sqrt{2}$, where $|H\rangle$ and $|V\rangle$ denotes the horizontal and vertical polarization states, respectively, and the subscript 1 and 2 denote the two output spatial modes. Extensive efforts are taken to ensure this source is robust against various vibration, temperature, and electromagnetic conditions (see SOM for details). After the satellite is launched, the source brightness and fidelity are tested by sampling ~1% of each path of the entangled photon pairs for on-satellite analysis (Fig. 2B). Under a pump power of ~30 mW, the source emits 5.9 million entangled photon pairs per second, with a state fidelity of $0.907\pm0.007$.

As the entangled photons propagate from the satellite through the atmosphere to the two ground stations at 500-2000 km apart, several effects contribute to channel loss, including beam diffraction, pointing error, atmospheric turbulence and absorption. As the entangled photons cannot be amplified as classical signals, a robust and efficient satellite-to-ground entanglement distribution places more stringent requirements on the link efficiency than conventional satellite-based classical communications. In particular, our work necessitates a satellite payload with two telescopes capable of establishing two independent satellite-to-ground quantum links simultaneously.

To optimize the link efficiency, we combine a narrow beam divergence with a high-bandwidth and high-precision acquiring, pointing, and tracking (APT) technique. The two entangled beams are sent out near-diffraction-limited far-field divergence of ~10 μrad by two Cassegrain telescopes with apertures of 300 mm and 180 mm (see Fig. 3A), which have been optimized to eliminate chromatic and spherical abbreviations at ~810 nm wavelength. The overall optical efficiencies of the two telescopes are 45%-55%. At Delingha, Lijiang and Nanshan station, the receiving telescopies have diameter of 1200 mm, 1800 mm, and 1200 mm, respectively. Our experiment has achieved entanglement distribution both between Delingha and Lijiang, and between Delingha and Nanshan (see SOM). Below we first describe the experiment involving Delingha and Lijiang.

We design cascaded multi-stage close-loop APT systems in both the transmitters (Fig. 3A) and receivers (Fig. 3B). The transmitters use green (~532 nm) beacon lasers, whereas the receivers send red (~671 nm) beacon lasers, pointing to each other with a divergence of ~1.2 mrad. The coarse pointing stages consist of a two-axis turntable or gimbal mirror, and wide field-of-view cameras, and achieve an accuracy better than 50 μrad. Further, the fine pointing stages with fast-steering mirrors and high-speed cameras, reliably lock the remote telescopes by a feedback closed-loop with a measured accuracy of 0.41 μrad in both x and y axis (Fig. 3C), much smaller than the beam divergence. For more technical details, see SOM. The APT systems are started when the satellite reaches 5° elevation angle, and the measurement begins at 10° elevation angle.

The relative motion of the satellite to the ground induces a drift of arrival time and polarization rotation seen by the receivers. We keep track of the relative motion between the transmitters and the receivers, and all the optical elements in the optical paths, to calculate the polarization rotation angle offset and phase shift. Using a motorized wave-plates combination (two quarter-wave plates and one half-wave plate) for dynamical polarization compensation, we can recover the polarization contrast to 80:1 (see SOM). Synchronization of two ground stations is done with a 100-KHz pulsed laser, sent from the satellite and in a good co-alignment with the signal photons. A synchronization jitter of 0.77 ns is obtained, which is used to tag the received signals and perform coincidence detection within a 2.5-ns time window. In addition to the temporal filtering, we place 20-nm bandwidth filters in the receiving telescope to reduce the background noise. In our experiment, depending on the position of the Moon, the background noise ranges from 500 to 2000 counts/s in each detector.

The satellite flies along a sun-synchronous orbit, and comes into both Delingha's and Lijiang's view once every night, starting at around 1:30AM Beijing time and lasting for a duration of ~275 seconds. Figure 4A plots the physical distances from the satellite to Delingha and Lijiang during one orbit, together the sum channel length of the two downlinks. Using a reference laser (see SOM) on the satellite, we measure in real time the overall two-downlink channel attenuation, which varies from 64 dB to 82 dB, as shown in Fig. 4B. A slight asymmetry is observed in the attenuation curve—when the satellite moves closer to Lijiang, the link efficiency is higher, which is because Lijiang station has a larger-size aperture telescope. We observe an average two-photon count rate of 1.1 Hz, with a signal to noise ratio of ~8:1. It is remarkable to note that, compared to the entanglement distribution method by direct transmission of the same two-photon source using the most common (with a loss of 0.2 dB/km) and best-performance (with a loss of 0.16 dB/km) commercial telecommunication fibers, respectively, the effective link efficiency of our satellite-based approach is 17 and 12 orders of magnitude higher (ref.强哥). The intrinsic physical loss limit of the silica optical fibers is estimated to be 0.095-0.13 dB/km (ref. OFT). Even with such perfect optical fibers if produced in the future, our satellite-based method is still 5-7 orders of magnitude more efficient. The efficiency of the fiber-based approach can in principle be enhanced by the use of quantum repeaters, however, the demonstrated quantum repeater nodes in the laboratories are still far from being practically applicable in realistic long-distance quantum communications.

The received photons are analyzed by a half-wave plate, a polarizing beam splitter, and a Pockel cell, then coupled into a multi-mode fiber and detected by single-photon detectors with ultra-low dark counts (<20 Hz). The Pockel cells are driven by high-voltage pulses rapidly switching between zero- and half-wave voltages, controlled by fast (4 Mbit/s) random numbers. Such a setting allows measurements of polarization at the basis of $\cos\theta|H\rangle + \sin\theta|V\rangle$. To verify whether the two photons, after passing the overall distance ranging from 1600 km to 2400 km, are still entangled, we analyze their polarizations in the $|H\rangle/|V\rangle$ and $|\pm\rangle = (|H\rangle \pm |V\rangle)/\sqrt{2}$ basis. We obtain 134 coincidence counts—raw data without subtracting background noise—during an effective time of 250s in satellite-orbit shadow time, where the data are plotted in Fig. 5B. We find that, in good agreement with the state $|\psi\rangle_{12}$, the $|H\rangle_1 |V\rangle_2$ and $|V\rangle_1 |H\rangle_2$ population dominates in the $|H\rangle / |V\rangle$ basis. Further, the coherence of the state is evident in Fig. 3B where the measured $|+\rangle_1 |+\rangle_2$ and $|-\rangle_1 |-\rangle_2$ counts dominate over $|+\rangle_1 |-\rangle_2$ and $|-\rangle_1 |+\rangle_2$, with a contrast of 16:1. From these measurements, we can estimate the state fidelity (defined as the wavefunction overlap of the experimentally obtained states with the ideal $|\psi\rangle_{12}$, ref. *26*) of the two photons distributed over 1203 km: $F \geq 0.87 \pm 0.09$, which is well above the threshold for both confirming the two-particle entanglement and violating Bell inequalities.

We use the distributed entangled photons for the Bell test using the Clauser-Horne-Shimony-Holt (CHSH)-type inequality (*27*), which is given by

$$S = \left| E(\varphi_1, \varphi_2) - E(\varphi_1, \varphi_2') + E(\varphi_1', \varphi_2) + E(\varphi_1', \varphi_2') \right| \leq 2,$$

where $E(\varphi_1, \varphi_2)$ is the joint correlation at the two remote locations with measurement angles of $\varphi_1$ and $\varphi_2$ respectively. The angles are randomly selected among (0, $\pi/8$), (0, $3\pi/8$), ($\pi/4$, $\pi/8$) and ($\pi/4$, $3\pi/8$) fast enough to close the locality (*28*) and freedom-of-choice loophole (Fig. 6A). We run 1167 trials of the Bell test during an effective time of 1059 s. The observed data in the four settings are summarized in Fig. 6B, from which we find $S = 2.37 \pm 0.09$, with a violation of the CHSH-Bell inequality $S \leq 2$ by 4 standard deviations. The result again confirms the non-local feature of entanglement and excludes the models of reality which sits on the notions of locality and realism, in a new space scale with thousands of kilometers.

In summary, we have demonstrated the distribution of two entangled photons from a satellite to two ground stations that are physically separated by 1203 km, and observed a survival of entanglement and violation of Bell inequality. The distributed entangled photons are readily useful for entanglement-based quantum key distribution (*7*), so far the only way demonstrated to establish secure keys between two distant locations with a separation of thousands of kilometers on the Earth without relying on trustful relay. Another immediate application is to exploit the distributed entanglement to perform a variant of quantum teleportation protocol (*29*) for remote preparation and control of quantum states, which can be a useful ingredient in distributed quantum networks. The developed satellite-based technology opens up a new avenue to both practical quantum communications and fundamental quantum optics experiments at distances inaccessible previously on the ground (*30*).

**Figure caption:**

**Figure 1: Overview of the experimental set-up of satellite-based entanglement distribution.** An illustration of the *Micius* satellite and the two ground stations in Delingha and Lijiang (see SOM for the third station in Nanshan). The satellite flies with a speed of 7.6 km/s along a sun-synchronous orbit at an altitude of ~500 km. The physical distance between Delingha and Lijiang ground station is 1203 km. Two independent transmitters equipped in the satellite sends the entangled two photons to the ground, enabled by high-precision APT.

**Figure 2: Schematic of spaceborn entangled-photon source and its in-orbit performance. (A)** The thickness of the PPKTP crystal is 15 mm. A pair of off-axis concave mirrors focus the pump laser in the center of the PPKTP. At the output of the Sagnac interferometer, two dichromatic mirrors (DMs) and long-pass filters are used

to separate the signal photons from the pump laser. Two additional electrically driven piezo steering mirrors (PIs), remotely controllable on the ground, are used for fine adjustment of the beam pointing for an optimal collection efficiency into the single-mode fibers. (**B**) The two-photon correlation curves measured by sampling 1% on each path of the entangled photons in the satellite. The count rate measured from the overall 0.01% sampling is about 590 Hz, from which we can estimate the source brightness of 5.9 MHz.

**Figure 3: The transmitters, receivers, and APT performance.** (**A**) The entangled photons beam (810 nm) is combined and co-aligned with a pulsed infrared laser (850 nm) for synchronization, and a green laser (532 nm) for tracking by three DMs and sent out from an 8× telescope. For polarization compensation, two motorized quarter-wave plates (QWPs) and a half-wave plate (HWP) are remotely controlled. An FSM and a two-axis turntable are used for closed-loop fine and coarse tracking, based on the 671-nm beacon laser images captured by the camera 1 and 2. (**B**) Schematic of the receiver at Delingha. The cooperating APT and polarization compensation systems are the same as those on the satellite. The tracking and synchronization lasers are separate from the signal photon and detected by single photon detectors. For polarization analysis along fast randomly switching bases, two QWPs, a HWP, a Pockel cell, and a PBS are used. (**C**) The APT systems starts tracking after the satellite reaches 5° elevation angle. The left panel is a 50s trace of the real-time images readout from the camera. Fine tracking accuracy of ~0.41 μrad is achieved in both x and y axis.

**Figure 4: Physical distances from satellite to two ground stations and the measured channel attenuation.** (**A**) A typical two-downlink from the satellite to Delingha and Lijiang that lasted for about 275 seconds in one orbit. The distance from the satellite to Delingha varies from 545 km to 1680 km. The distance from the satellite to Lijiang varies from 560 km to 1700 km. The overall length of the two-downlink channel varies from 1600 km to 2400 km. (**B**) The measured two-downlink channel attenuation in one orbit. The highest loss is ~82 dB at the sum distance of 2400 km, when the satellite just reached 10° elevation angle seen from Lijiang station.

As the telescope has a diameter of 1.8 m (the largest) and thus has a higher receiving efficiency than others stations, when the satellite flies over Lijiang at an elevation angle more than 15°, the channel loss remains relatively stable from 64dB to 68.5dB.

**Figure 5: Measurement of the received entangled photons after the two-downlink channel.** (**A**) Normalized two-photon coincidence counts in the measurement setting of $|H\rangle/|V\rangle$ basis. (**B**) Normalized counts in the diagonal $|\pm\rangle$ basis. Numbers in the brackets represent the raw coincidence counts of different measurement settings.

**Figure 6: Space-time diagram and Bell inequality violation.** (**A**) The top panel illustrates the space-time relationship among the entanglement generation point (S), the quantum random number generation points, R1 and R2, and the measurement results points, M1 and M2. The horizontal axis denotes the distances between the ground station and the satellite, which vary from 500 km to 1700 km. In our current experimental configuration, M1 and M2 are about 100 ns behind the light cone of S. The quantum random number generating rate is 5 kHz with an output delay below 200 ns. That is, the duration between R1 (R2) and M1 (M2) is in the range of 0.2 μs to 200.2 μs. Therefore, R1 (R2) and S are space-like separated, which implying the freedom-of-choice loophole is distinctly closed, under the additional assumption that all the possible hidden variables must originate together with the entangled particles. The bottom panel illustrates the relationship between two ground stations, which are 1203 km apart. Taking into account the orbit height of 500 km, the length difference of the two free-space channel does not exceed 944 km. Thus, the space-like criteria is satisfied between R1 and R2, R1 and M2, M1 and R2, M1 and M2. As a result, the locality loophole is addressed. (**B**) Correlation functions of a CHSH-type Bell's inequality for entanglement distribution. The measurement settings represent the measured polarization based of photons by Delingha and Lijiang respectively. Error bars are one standard deviation calculated from propagated Poissonian counting statistics of the raw photon detection events.

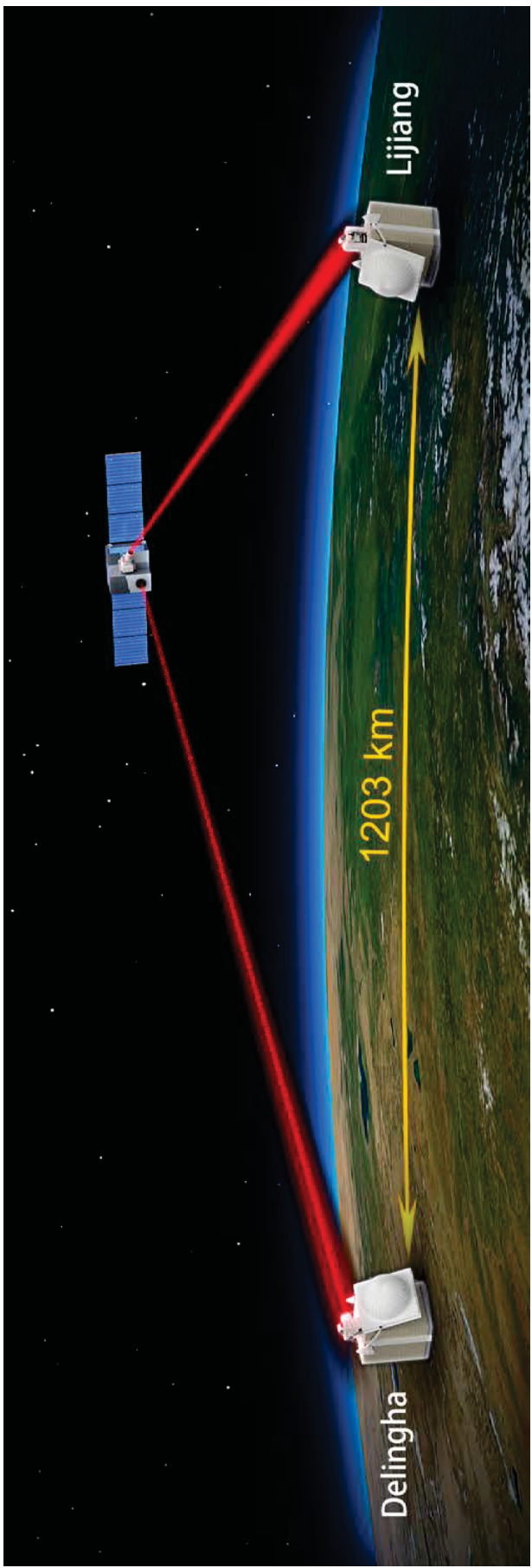

Fig. 1

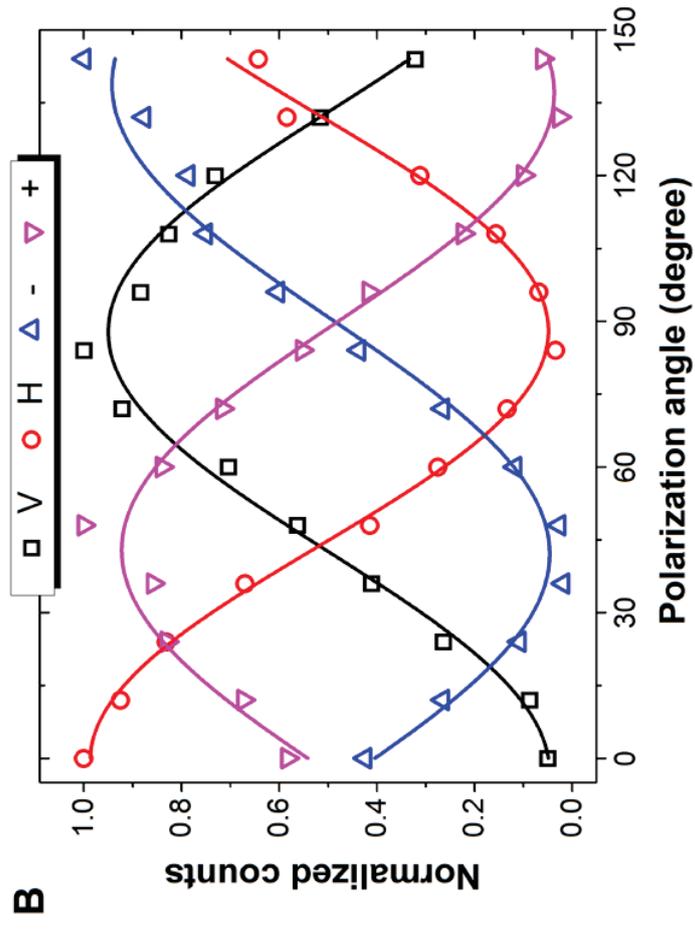

B

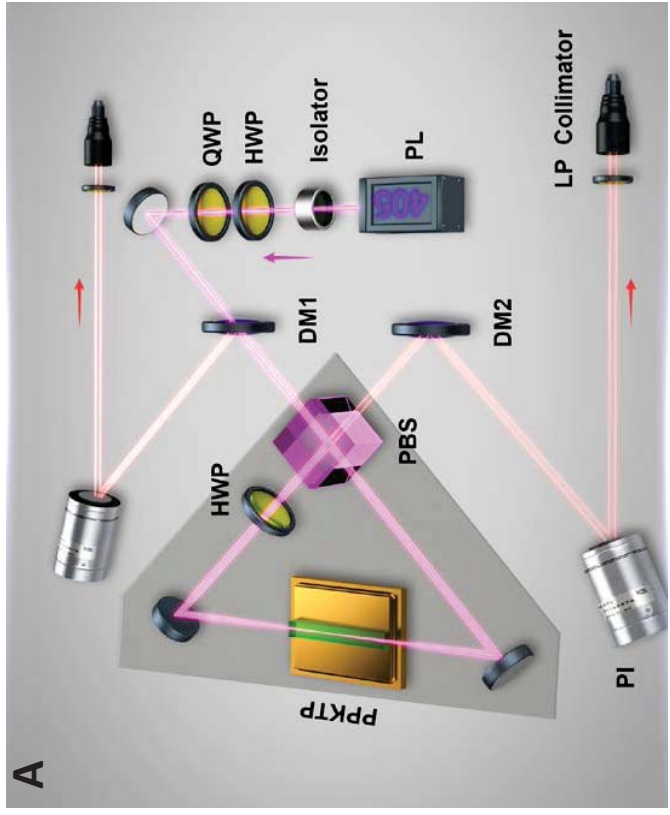

Fig. 2

Fig. 3

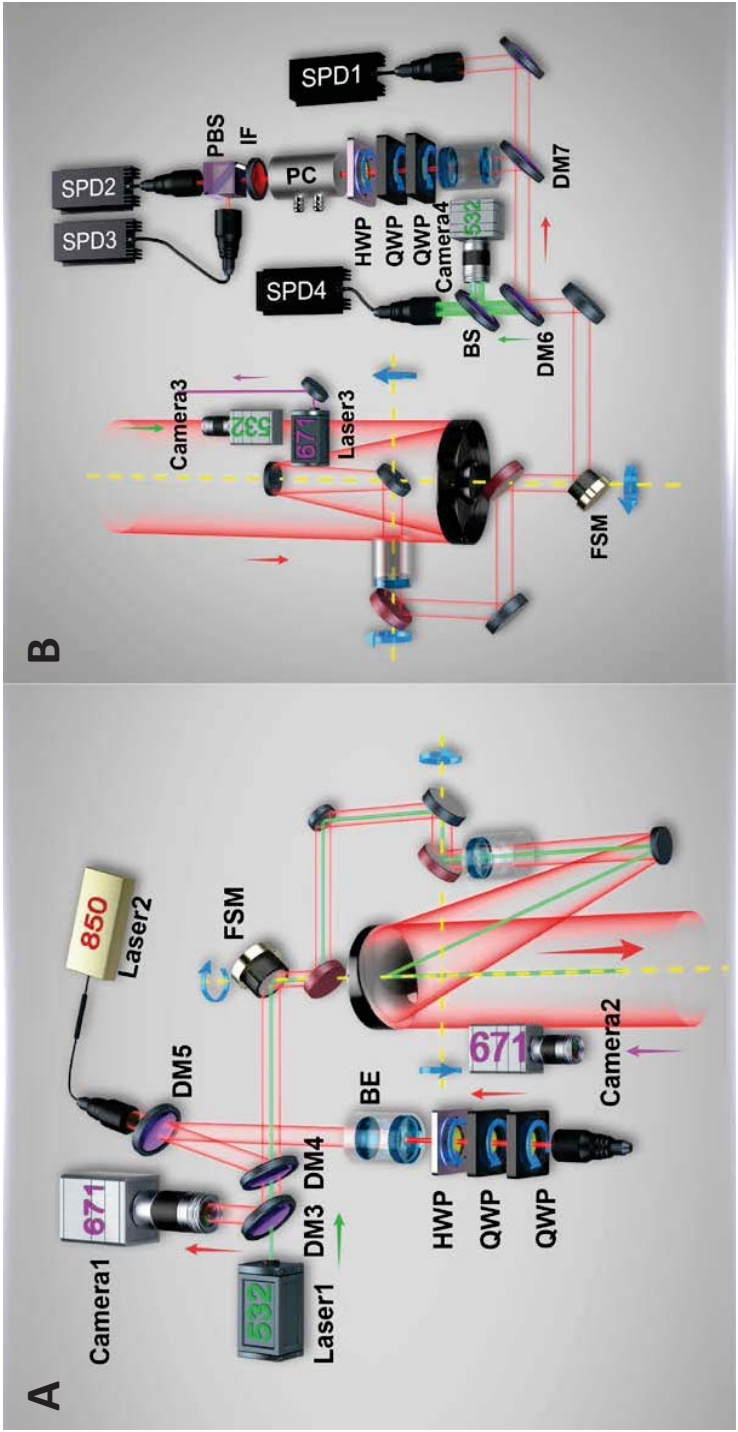

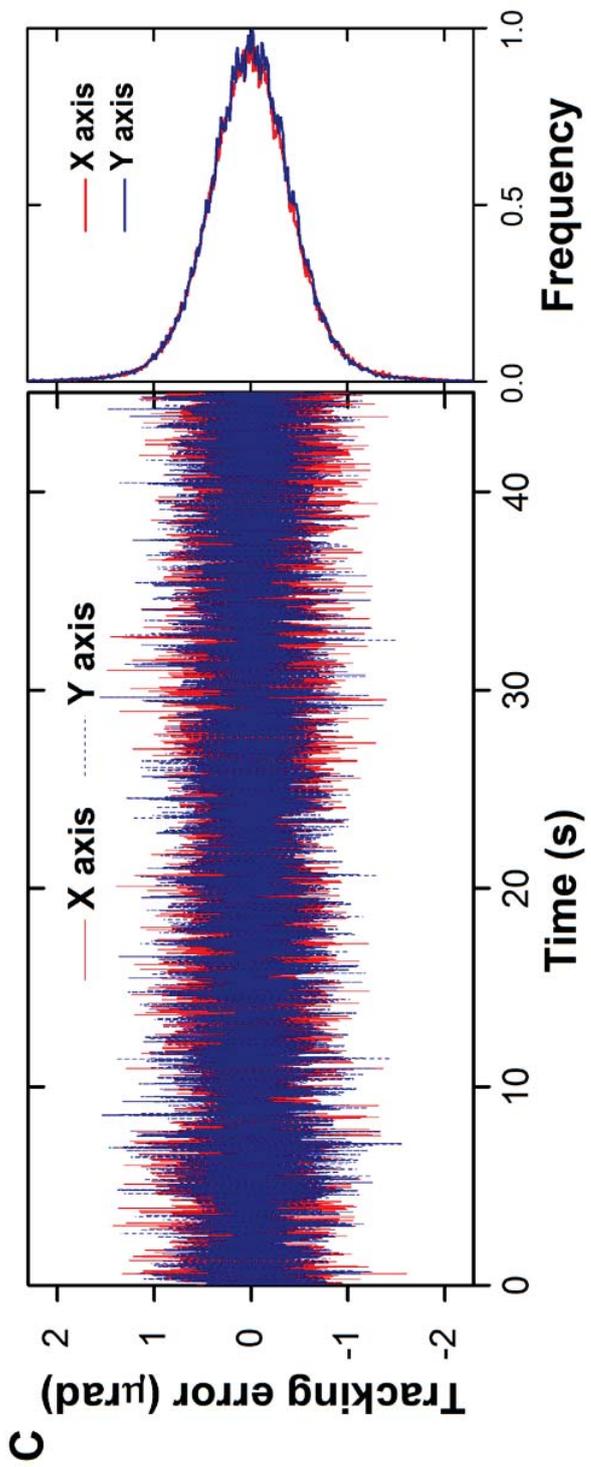

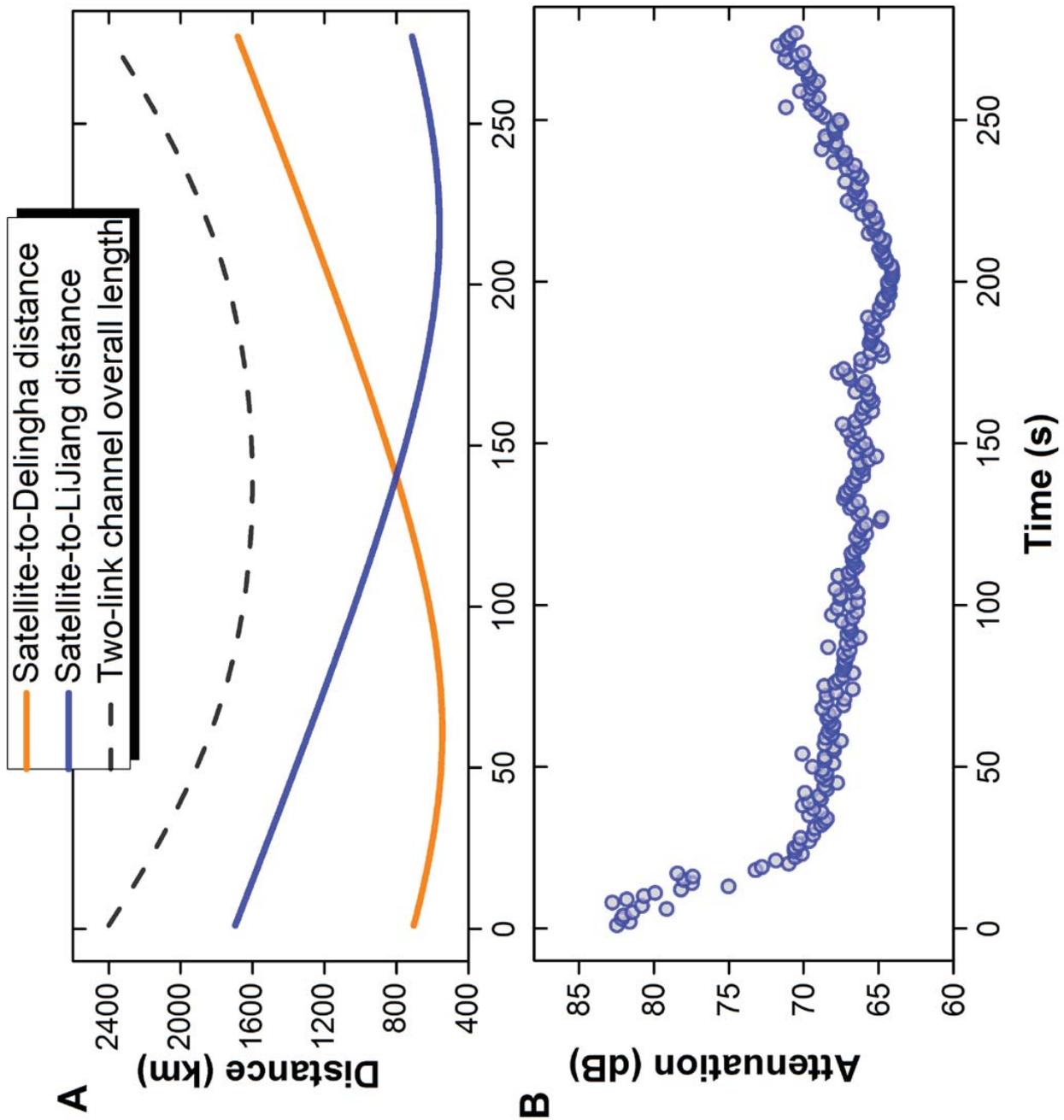

Fig. 4

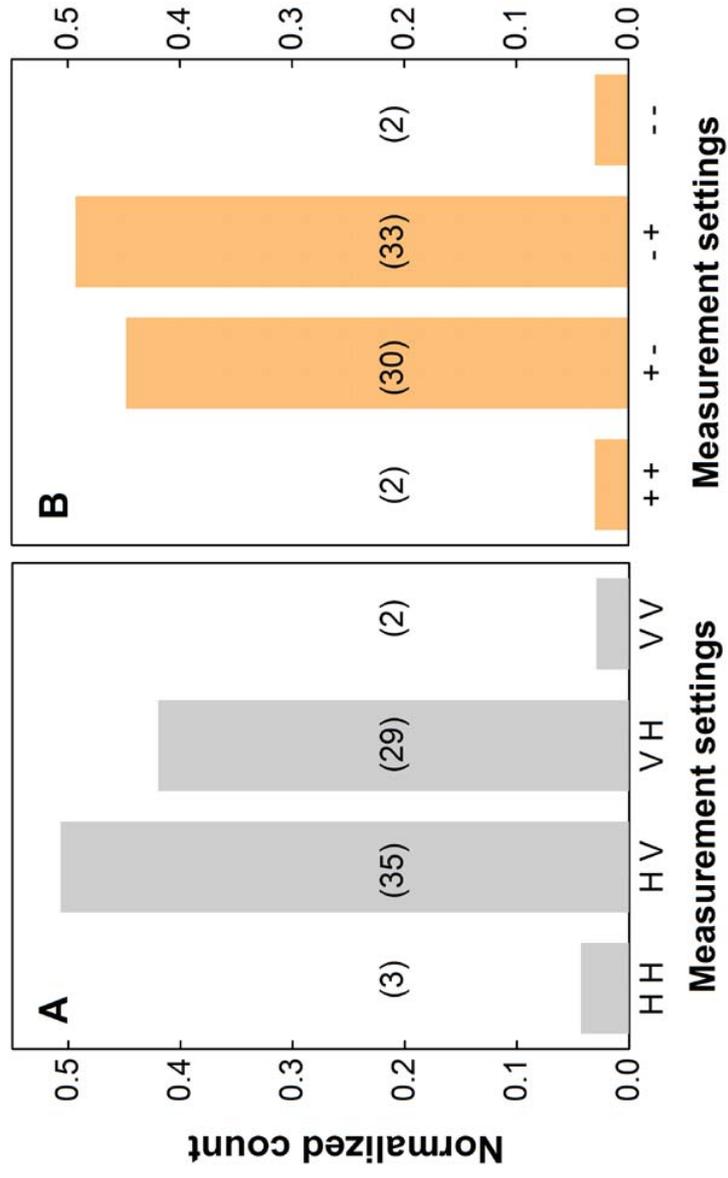

Fig. 5

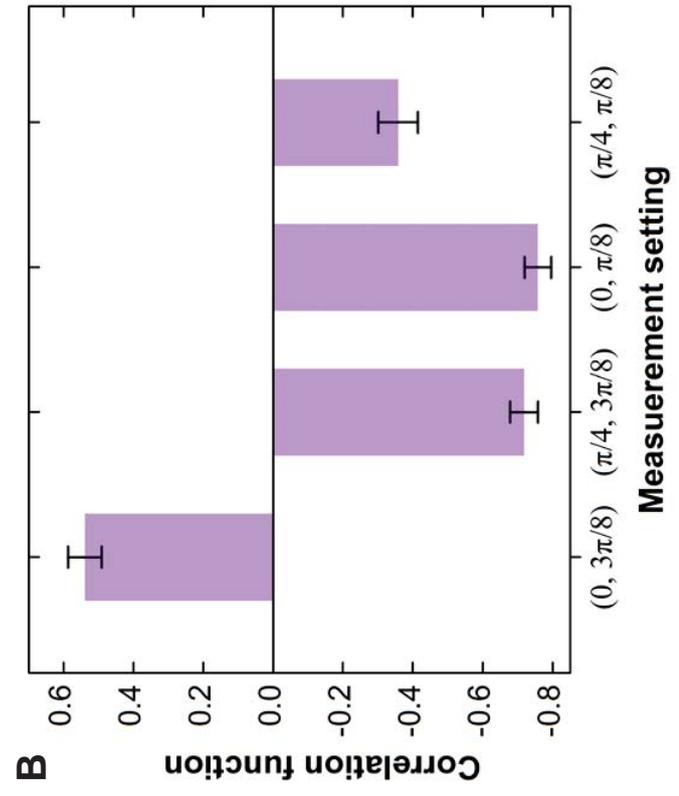

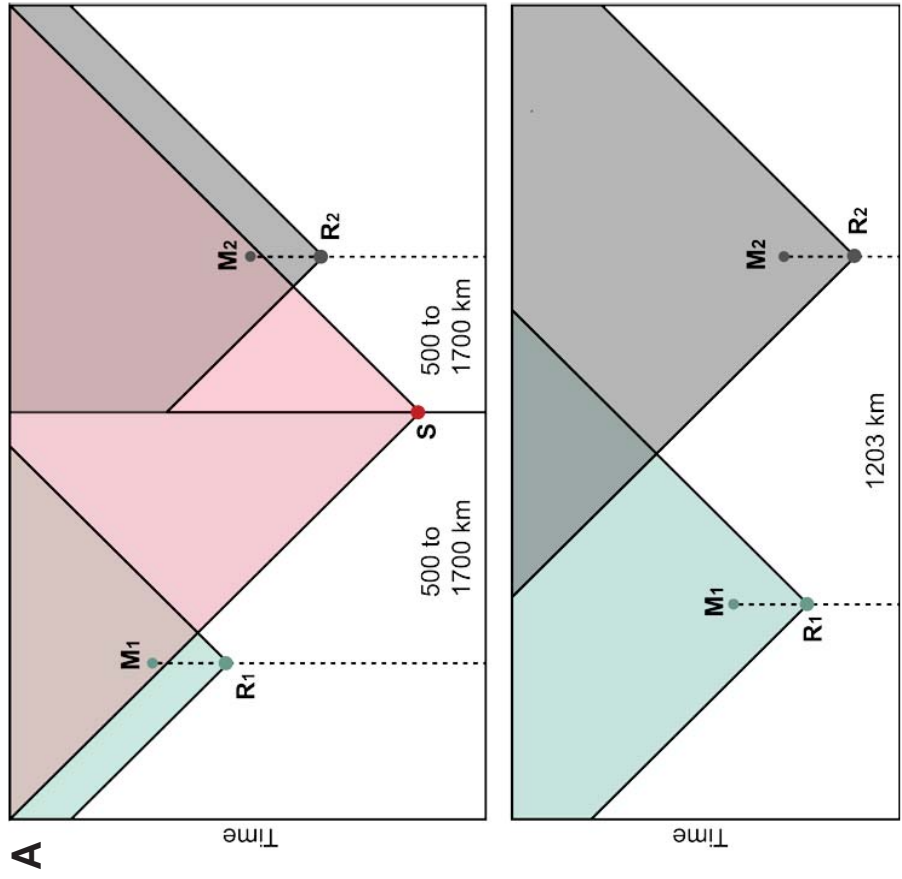

Fig. 6